\documentclass{osa-article}

\journal{osajournal}

\articletype{Research Article}

\usepackage{amsmath}
\usepackage{graphicx}
\usepackage{dcolumn}
\usepackage{bm}

\usepackage{bbold}

\newcommand{\half}{\ensuremath{\frac{1}{2}}}
\newcommand{\beqn}{\begin{equation}}
\newcommand{\eeqn}{\end{equation}}
\newcommand{\bse}{\begin{subequations}}
\newcommand{\ese}{\end{subequations}}
\newcommand{\del}{\partial}
\newcommand{\colvectwo}[2]{\begin{pmatrix} {#1}\\{#2}\end{pmatrix}}
\newcommand{\colvecthree}[3]{\begin{pmatrix}{#1}\\{#2}\\{#3}\end{pmatrix}}
\newcommand{\ket}[1]{\left|{#1}\right\rangle}
\newcommand{\bra}[1]{\left\langle {#1}\right|}
\newcommand{\bracket}[3]{\left\langle {#1}\left|{#2}\right|{#3}\right\rangle}
\newcommand{\braket}[2]{\left\langle {#1} \middle| {#2} \right\rangle}
\newcommand{\parder}[2]{\frac{\del {#1}}{\del {#2}}}

\begin{document}

\title{Nonlinear Polarization Transfer and Control of Two Laser Beams Overlapping in a Uniform Nonlinear Medium}

\author{Eugene Kur,\authormark{1,*} Malcolm Lazarow,\authormark{1} Jonathan S. Wurtele,\authormark{1} and Pierre Michel\authormark{2}}

\address{\authormark{1}Department of Physics, University of California, Berkeley, California 94720, USA\\
\authormark{2}Lawrence Livermore National Laboratory, Livermore, California 94551, USA}

\email{\authormark{*}k-gene@berkeley.edu} 



\begin{abstract}
A scheme for polarization control using two laser beams in a non-linear optical medium is studied using both co- and counter-propagating beam geometries. In particular, we show that under certain conditions it is possible for two laser beams to exchange their polarization states. A model accounting for a more realistic, 2D propagation geometry is presented. The 2D model produces drastically different results (compared to the 1D propagation geometry), creating difficulties for implementing polarization control in a realistic setting. A proposal for overcoming these difficulties by reducing the non-linear optical medium to a thin slab is presented.
\end{abstract}

\section{Introduction}
Controlling the polarization state of a laser beam is of particular interest in optical applications (see e.g. \cite{millot_2014,assemat_2010} and references therein). Traditionally such control is accomplished by propagating the beam through a material whose index of refraction induces the desired polarization change. Such a scheme can be passive (the material already possesses the desired index of refraction) or active, where the desired index of refraction is obtained by application of external electric or magnetic fields.\cite{boyd_book}

Recently, interest has arisen in a new type of active control scheme using a second laser beam together with a nonlinear medium \cite{michel_PRL2014,michel_PRX2020}. When the beam of interest overlaps with the secondary laser, the two form a beat wave. The nonlinear medium has a refractive index of the form $n=n_0+n_2I$ ($I$ being the intensity of the laser field) so the beat wave produces a modulation of the index of refraction, allowing the nonlinear medium to facilitate polarization transfer between the laser beams. When the nonlinear medium is a plasma, this two-beam style of polarization control allows for the manipulation of intense laser beams without the need to worry about optics damage. The original theory \cite{michel_PRL2014} as well as subsequent experiments \cite{turnbull_PRL2016,goyon_PRL2017} worked in the regime where the second (auxiliary) laser beam was much more intense than the beam of interest (the linear regime), so a 1D model of beam propagation was sufficient to describe the polarization behavior. A subsequent analysis showed that polarization control is also possible in the nonlinear regime (when the beams have similar intensities) by crossing the beams at $90^\circ$, allowing manipulation of a more intense primary beam using a less intense auxiliary beam.\cite{michel_PRX2020}

In this paper, we extend the analysis of the nonlinear regime and unveil a new arrangement where the beams can either periodically exchange their polarization states, or periodically “flip” to orthogonal polarization states. The former is achieved for two beams co-propagating at a small angle, and the latter for two counter-propagating beams. The two beams must have equal intensities in either case. We show that a realistic 2D propagation geometry poses several challenges for implementing a polarization control scheme, but propose a geometrical arrangement based on a thin nonlinear medium which allows us to recover the 1D interaction. This scheme could have a wide range of applications in areas such as optical computing or cryptography. It can also allow for light manipulation at extreme fluences and ultrafast time-scales when the nonlinear medium is a plasma or partially ionized gas.

The organization of the paper is as follows. We first review polarization dynamics in a 1D propagation geometry in Section~\ref{sec:1D}. We find that two beams of equal intensity and wavelength undergo periodic evolution of their polarizations, with either a ``swap" at each half-period when the beams are co-propagating (see Sec.~\ref{subsec:coprop}) or a ``flip" at each half-period when the beams are counter-propagating (see Sec.~\ref{subsec:counterprop}). In Section~\ref{sec:2D}, we extend the 2D intensity model of Ref.~\cite{mckinstrie_1996} to account for polarization dynamics. Using this new model, we show how the 2D propagation geometry dramatically changes the polarization dynamics from the 1D case. In Section~\ref{sec:ThinPlasma} we present our method of recovering the 1D polarization dynamics in a 2D geometry by using a thin slab of nonlinear medium. Some potential applications of these concepts to optical computing and cryptography are presented in the conclusion (Sec.~\ref{sec:conclusion}).

\section{Nonlinear Polarization Swapping and Control in 1D\label{sec:1D}}

We consider two propagating laser beams in the eikonal approximation producing a total electric field $\tilde{\bm{E}}=\half\bm{E}_0(\bm{r},t)e^{i\psi_0}+\half\bm{E}_1(\bm{r},t)e^{i\psi_1}+\text{c.c.}$, where $\psi_i=\bm{k}_i\cdot\bm{r}-\omega_it$ and we place a tilde on rapidly-varying fields to distinguish them from the slowly-varying envelopes. The beams propagate through a nonlinear medium with index of refraction $n=n_0+n_2I$, where $n_0$ is the linear refractive index, $n_2$ specifies the strength of the nonlinearity ($n_2$ is independent of the laser intensity), and $I=n_0c\langle\tilde{\bm{E}^2}\rangle/4\pi$ is the intensity \cite{boyd_book}. We will assume $n_0$ and $n_2$ are uniform and constant throughout the paper. Averaging over the fast oscillations gives a total index of refraction 

\beqn\label{eq:index_modulation}
n=n_0+\frac{n_0n_2c}{4\pi}\left[\half\bm{E}_0\cdot\bm{E}_1^*e^{i\psi_b}+\text{c.c}\right],
\eeqn where $\psi_b=\psi_0-\psi_1$ (we also define $\omega_b=\omega_0-\omega_1$ and $\bm{k}_b=\bm{k}_0-\bm{k}_1$). We focus on the case where there is only polarization dynamics and no energy is exchanged between the beams. Thus $\omega_0=\omega_1$ and there is no phase delay between the index of refraction modulation and the beat wave produced by the laser beams (c.f. Ref.~\cite{michel_PRX2020}). Extending the formalism to account for energy exchange is straightforward (see e.g. Ref.~\cite{michel_PRL2014}).

Substituting the electric field and index of refraction into the nonlinear wave equation $(\nabla^2-\frac{n^2}{c^2}\del_t^2)\bm{E}=0$, using the eikonal assumption ($|\nabla \bm{E}_i|\ll k_i |\bm{E}_i|$, $|\del_t\bm{E}_i|\ll\omega_i|\bm{E}_i|$), and collecting terms oscillating at $\psi_0,\psi_1$ gives the coupled amplitude equations
\bse\label{eqs:full_amplitude}
\begin{align}
	i(\bm{k}_0\cdot\nabla)\bm{E}_0 &= -\frac{n_2k_0^2c}{8\pi}(\bm{E}_0\cdot\bm{E}_1^*)\bm{E}_1,\label{eq:full_amplitude_0}\\
	i(\bm{k}_1\cdot\nabla)\bm{E}_1 &= -\frac{n_2k_1^2c}{8\pi}(\bm{E}_1\cdot\bm{E}_0^*)\bm{E}_0,\label{eq:full_amplitude_1}
\end{align}
\ese where we have assumed a steady-state solution is reached ($\del_t\bm{E}_i=0\implies \bm{E}_i=\bm{E}_i(\bm{r})$).

We assume a 1D model with the beams propagating along $z$ so $\bm{E}_i=\bm{E}_i(z)$, $\bm{k}_0=k_0\hat{\bm{z}}$, and $\bm{k}_1=\pm k_1\hat{\bm{z}}$ with $+$ for co-propagation and $-$ for counter-propagation. Selecting two basis vectors $\hat{\bm{x}}$ and $\hat{\bm{y}}$ in the plane orthogonal to propagation, we employ the (normalized) Jones vector notation 
\beqn\label{eq:Jones_vector}
\ket{\mathcal{E}_i}=\frac{1}{\sqrt{|E_{x,i}|^2+|E_{y,i}|^2}}\colvectwo{E_{x,i}}{E_{y,i}},\,\bra{\mathcal{E}_i}=\frac{1}{\sqrt{|E_{x,i}|^2+|E_{y,i}|^2}}\begin{pmatrix}
	E_{x,i}^* & E_{y,i}^*
\end{pmatrix}.
\eeqn Note $\braket{\mathcal{E}_i}{\mathcal{E}_i}=1$. 

We will use equal intensity beams ($I_0=I_1=I$) as this gives the simplest method for polarization control. Methods involving unequal beam intensities can also be developed by a straightforward extension of our formalism (see e.g. Ref.~\cite{turnbull_PRL2016} for a limiting case). With equal beam intensities (and equal beam frequencies $k_0=k_1=k$) we can use a single normalized $z$-coordinate 
\beqn\label{eq:normalized_z}
\zeta=\frac{n_2I}{n_0}kz,
\eeqn
resulting in the simplified amplitude equations (see also Refs.~\cite{david_1990,pitois_2005})
\bse\label{eqs:simple_amplitude}
\begin{align}
\parder{}{\zeta}\ket{\mathcal{E}_0} &=i\ket{\mathcal{E}_1}\braket{\mathcal{E}_1}{\mathcal{E}_0}\label{eq:simple_amplitude_0},\\
\parder{}{\zeta}\ket{\mathcal{E}_1} &=\pm i\ket{\mathcal{E}_0}\braket{\mathcal{E}_0}{\mathcal{E}_1}.\label{eq:simple_amplitude_1}
\end{align}
\ese We now look at the implications of these equations for polarization control of co- and counter-propagating beams.

\subsection{Co-propagation\label{subsec:coprop}}
Equal frequency, equal intensity beams in the co-propagating geometry exhibit polarization swapping as the two beams periodically exchange their initial polarizations. We demonstrate this by explicitly solving Eqs.~\eqref{eqs:simple_amplitude}. 

We begin by differentiating Eq.~\eqref{eq:simple_amplitude_0} and substituting in Eq.~\eqref{eq:simple_amplitude_1} (using the $+$ sign for co-propagation). This gives the second-order equation 
\beqn\label{eq:coprop_2ndorder_0}
\frac{\del^2}{\del \zeta^2}\ket{\mathcal{E}_0}=-\left|\braket{\mathcal{E}_0}{\mathcal{E}_1}\right|^2\ket{\mathcal{E}_0}.
\eeqn Furthermore, $\braket{\mathcal{E}_0}{\mathcal{E}_1}$ is conserved:
\beqn
\parder{}{\zeta}\braket{\mathcal{E}_0}{\mathcal{E}_1}=-i\braket{\mathcal{E}_0}{\mathcal{E}_1}+i\braket{\mathcal{E}_0}{\mathcal{E}_1}=0.
\eeqn The solution to Eq.~\eqref{eq:coprop_2ndorder_0} can thus be written
\beqn\label{eq:coprop_soln}
\ket{\mathcal{E}_0(\zeta)} = \ket{\mathcal{E}_0(0)}\cos\left(\left|\braket{\mathcal{E}_0}{\mathcal{E}_1}\right| \zeta\right) + i\ket{\mathcal{E}_1(0)}\frac{\braket{\mathcal{E}_1}{\mathcal{E}_0}}{\left|\braket{\mathcal{E}_0}{\mathcal{E}_1}\right|}\sin\left(\left|\braket{\mathcal{E}_0}{\mathcal{E}_1}\right| \zeta\right),
\eeqn with a similar solution for $\ket{\mathcal{E}_1(\zeta)}$ (obtained by exchanging the labels $0$ and $1$).

Polarizations of the two beams have periodic evolution with period
\beqn\label{eq:swap_period}
\zeta_{\pi}=\frac{\pi}{\left|\braket{\mathcal{E}_0}{\mathcal{E}_1}\right|}.
\eeqn This period brings the polarizations back to their initial state up to an overall phase. Halfway through a period, the polarization of beam 0 becomes beam 1's initial polarization up to an overall phase 
\beqn\label{eq:beam_swap}
\ket{\mathcal{E}_0(\zeta_{\pi}/2)}=i\ket{\mathcal{E}_1(0)}\frac{\braket{\mathcal{E}_1}{\mathcal{E}_0}}{\left|\braket{\mathcal{E}_0}{\mathcal{E}_1}\right|},
\eeqn and similarly for beam 1. Thus the beams swap their polarizations after a physical distance 
\beqn\label{eq:z_swap}
z_\text{swap}=\frac{n_0}{n_2Ik}\frac{\pi}{2\left|\braket{\mathcal{E}_0}{\mathcal{E}_1}\right|}.
\eeqn

\begin{figure}
\centering
	\includegraphics[scale=0.7]{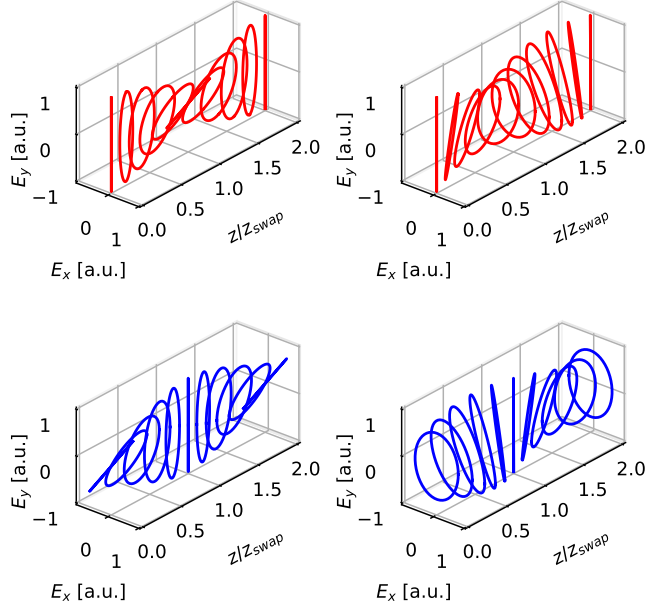}
	\caption{\label{fig:1Dco_ellipses} Polarization ellipses displaying the evolution of two co-propagating beams (both beams propagate in $z$-direction) obtained by solving Eqs.\eqref{eqs:simple_amplitude}. Evolution is periodic, with the two beams swapping their initial polarization after half a period. Beam 0 (top, red) is initially s-polarized while beam 1 (bottom, blue) is initially either linearly polarized at $45^\circ$ (left) or right-circularly polarized (right).}
\end{figure}

We demonstrate two versions of this solution in Fig.~\ref{fig:1Dco_ellipses}. The polarizations of the two beams are displayed over a single period. Beam 0 starts $y$-polarized in both cases while beam 1 starts either polarized at $45^\circ$ to the $x$-axis or is right-circularly polarized. Halfway through a period, the polarization of beam 0 matches the initial polarization of beam 1 (and vice versa). We are thus able to change the polarization state of beam 0 into any state we wish by selecting the corresponding initial polarization for beam 1. This allows us to control the polarization of a beam using a second beam of equal intensity.

The 1D swapping solution has a simple description on the Poincare sphere. In our notation the (unit) Stokes vector $\hat{\bm{J}}$ corresponding to a Jones vector $\ket{\mathcal{E}}$ is (see e.g. Ref.~\cite{gordon_2000})
\beqn\label{eq:Stokes_vector}
\hat{\bm{J}}=\frac{1}{|E_x|^2+|E_y|^2}\colvecthree{|E_x|^2-|E_y|^2}{2\mathcal{R}(E_xE_y^*)}{-2\mathcal{I}(E_xE_y^*)}=\colvecthree{\bracket{\mathcal{E}}{\sigma_z}{\mathcal{E}}}{\bracket{\mathcal{E}}{\sigma_x}{\mathcal{E}}}{\bracket{\mathcal{E}}{\sigma_y}{\mathcal{E}}},
\eeqn
where $\sigma_i$ are the Pauli matrices. The Stokes vector represents a polarization state as a point on the surface of the unit sphere in $\mathbb{R}^3$. This is the Poincare sphere and each beam's polarization traces out a curve on this sphere as the beam propagates. Evolution on the Poincare sphere is governed by the Stokes vector equations (equivalent to Eqs.~\eqref{eqs:simple_amplitude}) 
\bse\label{eqs:simple_stokes}
\begin{align}
\parder{}{\zeta}\hat{\bm{J}}_0&=\hat{\bm{J}}_0\times\hat{\bm{J}_1},\label{eq:simple_stokes_0}\\
\parder{}{\zeta}\hat{\bm{J}}_1&=\hat{\bm{J}}_1\times\hat{\bm{J}_0}.\label{eq:simple_stokes_1}
\end{align}
\ese We will plot the polarization trajectories of both beams on a single sphere for clarity.

\begin{figure}
\centering
	\includegraphics[scale=0.7]{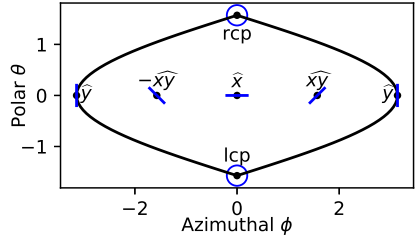}
	\caption{\label{fig:sphere_intro} The sinusoidal projection of the Stokes's sphere. Special polarization states are labeled and their polarization ellipses displayed. Linear polarizations lie along the equator, circular polarizations are at the poles (right-circular polarization at the north pole, left-circular polarization at the south pole), while the rest of the sphere has elliptical polarization.}
\end{figure}

To plot the Poincare sphere in 2D we use the sinusoidal projection (see e.g. Ref.~\cite{snyder_1987})
\bse\label{eqs:sinusoidal_proj}
\begin{align}
x &=\phi\cos\theta=\sqrt{1-\left(\frac{2\mathcal{I}(E_xE_y^*)}{|E_x|^2+|E_y|^2}\right)^2}\tan^{-1}\left(\frac{2\mathcal{R}(E_xE_y^*)}{|E_x|^2-|E_y|^2}\right),\\
y &=\theta=\frac{\pi}{2}-\cos^{-1}\left(\frac{2\mathcal{I}(E_xE_y^*)}{|E_x|^2+|E_y|^2}\right),
\end{align}
\ese where $\phi\in[-\pi,\pi]$ is the azimuthal angle on the sphere (measured from the positive $x$-axis) and $\theta\in[-\pi/2,\pi/2]$ is the polar angle on the sphere (measured from the equator). The sinusoidal projection of the Poincare sphere with a few sample polarization states is shown in figure~\ref{fig:sphere_intro}. Linear polarizations lie along the equator while the north and south pole represent right-handed and left-handed circular polarizations, respectively.

\begin{figure}
\centering
	\includegraphics[scale=0.7]{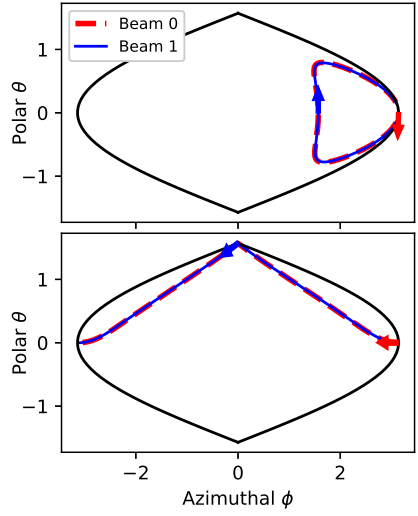}
	\caption{\label{fig:1Dco_sphere} Polarization trajectories of two co-propagating beams displayed on the Stokes's sphere. Arrows indicate initial state and direction of evolution. The two beams trace out a circle, always diametrically opposite to each other. Thus the beams swap polarizations after half a period. Beam 0 (thick red dashed line) is initially s-polarized while beam 1 (blue solid line) is initially either linearly polarized at $45^\circ$ (top) or right-circularly polarized (bottom).}
\end{figure}

Figure~\ref{fig:1Dco_sphere} illustrates the co-propagating solutions on the projected Poincare sphere. The trajectories are coincident circles, with the two beams remaining diametrically opposite each other. A half period of evolution thus exchanges the polarization states of the two beams, demonstrating the swapping behavior noted in Eq.~\eqref{eq:beam_swap}. The coincident circles are a direct consequence of Eqs.~\eqref{eqs:simple_stokes}. Adding the two equations shows $\hat{\bm{J}}_0+\hat{\bm{J}}_1$ is conserved. Thus the two unit vectors must rotate around their angle bisector, staying diametrically opposite on a circular trajectory.

\subsection{Counter-propagation\label{subsec:counterprop}}

We can similarly analyze the counter-propagating geometry. Here equal frequency, equal intensity beams exhibit polarization ``flipping". Evolution is again periodic and at the half-period each beam's polarization state becomes orthogonal to the other beam's initial polarization. We demonstrate this phenomenon by again solving Eqs.~\eqref{eqs:simple_amplitude} explicitly.

As before we differentiate Eq.~\eqref{eq:simple_amplitude_0} and substitute into Eq.~\eqref{eq:simple_amplitude_1} (using the $-$ sign for counter-propagation). This gives the second-order equation
\beqn\label{eq:counterprop_2ndorder_0}
\frac{\del^2}{\del\zeta^2}\ket{\mathcal{E}_0}=\left|\braket{\mathcal{E}_0}{\mathcal{E}_1}\right|^2\ket{\mathcal{E}_0}+2i\parder{}{z}\ket{\mathcal{E}_0},
\eeqn where we have re-substituted Eq.~\eqref{eq:simple_amplitude_0} in place of $\ket{\mathcal{E}_1}\braket{\mathcal{E}_1}{\mathcal{E}_0}$. In the counter-propagating geometry only the magnitude $\left|\braket{\mathcal{E}_0}{\mathcal{E}_1}\right|$ is conserved:
\beqn
\parder{}{\zeta}\left|\braket{\mathcal{E}_0}{\mathcal{E}_1}\right|^2 = 2\mathcal{R}\left[\braket{\mathcal{E}_1}{\mathcal{E}_0}\parder{}{\zeta}\braket{\mathcal{E}_0}{\mathcal{E}_1}\right]
= -4\mathcal{R}\left[i\left|\braket{\mathcal{E}_0}{\mathcal{E}_1}\right|^2\right]=0.
\eeqn This is sufficient to give Eq.~\eqref{eq:counterprop_2ndorder_0} constant coefficients and thus the solution
\beqn\label{eq:counterprop_soln}
\ket{\mathcal{E}_0(\zeta)} = e^{i\zeta}\left[\ket{\mathcal{E}_0(0)}\cos\left(\pi\zeta/\zeta_\pi\right)
+ i   \frac{\ket{\mathcal{E}_1(0)}\braket{\mathcal{E}_1(0)}{\mathcal{E}_0(0)}-\ket{\mathcal{E}_0(0)}}{\sqrt{1-\left|\braket{\mathcal{E}_0}{\mathcal{E}_1}\right|^2}}\sin\left(\pi\zeta/\zeta_\pi\right)\right],
\eeqn 
where $\zeta_\pi=\pi/\sqrt{1-\left|\braket{\mathcal{E}_0}{\mathcal{E}_1}\right|^2}$ is the period of the polarization evolution (c.f. Eq.~\eqref{eq:swap_period}). A similar solution for $\ket{\mathcal{E}_1(\zeta)}$ is obtained by exchanging the labels $0$ and $1$ and changing $\zeta\to-\zeta$ (on the right-hand side of the equation). Note that whereas $\ket{\mathcal{E}_0(0)}$ refers to the initial state of beam 0, $\ket{\mathcal{E}_1(0)}$ refers to the \emph{final} state of beam 1 due to the counter-propagating geometry. We can express Eq.~\eqref{eq:counterprop_soln} in terms of the initial state of beam 1 $\ket{\mathcal{E}_1(\zeta_f)}$ by using 
\beqn
\ket{\mathcal{E}_1(0)} = e^{i\zeta_f}\left[\mathbb{1} \cos\left(\pi\zeta_f/\zeta_\pi\right)
-i\frac{\ket{\mathcal{E}_0(0)}\bra{\mathcal{E}_0(0)}-\mathbb{1}}{\sqrt{1-\left|\braket{\mathcal{E}_0}{\mathcal{E}_1}\right|^2}}\sin\left(\pi\zeta_f/\zeta_\pi\right)\right]^{-1}\ket{\mathcal{E}_1(\zeta_f)},
\eeqn
where $\mathbb{1}$ is the $2\times 2$ identity matrix and prior knowledge of the constant $\left|\braket{\mathcal{E}_0(0)}{\mathcal{E}_1(0)}\right|^2$ is required to evaluate this expression. Note that when $\zeta_f=\zeta_\pi/2$ the matrix in brackets becomes singular and so non-invertible. In this case the final state of beam 1 $\ket{\mathcal{E}_1(0)}$ cannot be uniquely identified from the initial state $\ket{\mathcal{E}_1(\zeta_f)}$. This phenomenon, referred to as multi-stability, is explored further in Refs.~\cite{lytel_1984,kaplan_1985,gauthier_1990}.

Halfway through a period ($\zeta=\zeta_\pi/2$), the polarization of beam 0 becomes orthogonal to beam 1's final polarization:
\beqn\label{eq:beam_flip}
\braket{\mathcal{E}_1(0)}{\mathcal{E}_0(\zeta_{\pi}/2)}=ie^{i\zeta_\pi/2}\frac{\braket{\mathcal{E}_1(0)}{\mathcal{E}_0(0)}-\braket{\mathcal{E}_1(0)}{\mathcal{E}_0(0)}}{\sqrt{1-\left|\braket{\mathcal{E}_0}{\mathcal{E}_1}\right|^2}}=0.
\eeqn A similar result holds for beam 1. Thus the beams ``flip" their polarizations (switch to a polarization orthogonal to the other beam's) after a physical distance
\beqn
z_\text{flip}=\frac{n_0}{n_2Ik}\frac{\pi}{2\sqrt{1-\left|\braket{\mathcal{E}_0}{\mathcal{E}_1}\right|^2}}.
\eeqn

\begin{figure}
\centering
	\includegraphics[scale=0.7]{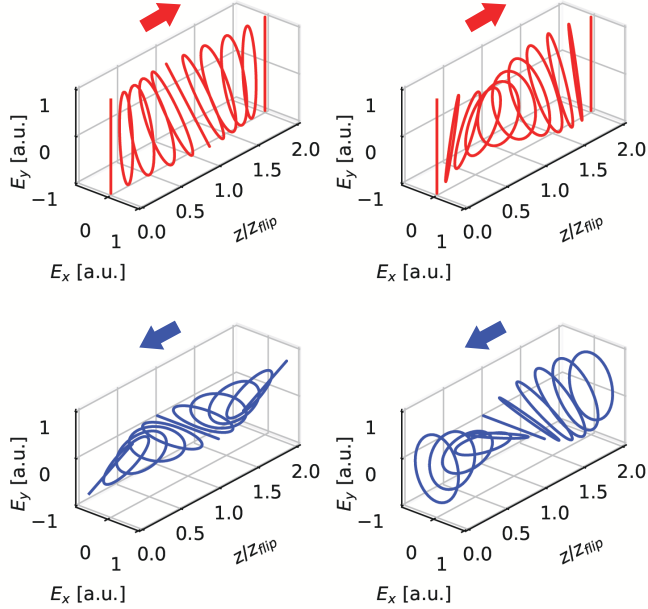}
	\caption{\label{fig:1Dcntr_ellipses} Polarization ellipses displaying the evolution of two counter-propagating beams, obtained by solving Eqs.~\eqref{eqs:simple_amplitude}. Beam 0 (top, red) propagates in the $z$-direction while beam 1 (bottom, blue) propagates in the $-z$-direction as indicated by the arrows. Evolution is periodic; at half a period each beam reaches a polarization orthogonal to the other beam's initial polarization. Beam 0 (top, red) is initially s-polarized while beam 1 (bottom, blue) is initially either linearly polarized at $45^\circ$ (left) or left-circularly polarized (right).}
\end{figure}

We demonstrate two versions of this solution in Fig.~\ref{fig:1Dcntr_ellipses}. The polarizations of the two beams are displayed over a single period. Beam 0 starts $y$-polarized in both cases while beam 1 starts either polarized at $45^\circ$ to the $x$-axis or is left-circularly polarized. By using a whole period, we need not distinguish between the initial and final polarization states of beam 1, simplifying our discussion. Halfway through a period, the polarization of beam 0 becomes orthogonal to the initial state of beam 1 (and vice versa). We are thus able to change the polarization state of beam 0 into any state we wish be selecting a corresponding orthogonal state for the initial polarization of beam 1. This allows us to control the polarization of a beam using a second counter-propagating beam of equal intensity.

As with the co-propagating swapping solution, the 1D counter-propagating flipping solution has a simple description on the Poincare sphere. The Stokes vector is defined as before, but now beam 1 is counter-propagating. The Stokes vector of beam 1 thus describes its polarization state with respect to beam 0's orientation. Thus if beam 1 is left-circularly polarized, for example, its Stokes vector would be $\hat{\bm{J}}_1=\begin{pmatrix} 0 & 0 & 1\end{pmatrix}$ corresponding to the north pole of the Poincare sphere (normally representing right-circular polarization). This is because when beam 1 is left-circularly polarized relative to its own direction of propagation, it is right-circularly polarized from beam 0's perspective. This type of orientation reversal of the Stokes vector makes the solution on the Poincare sphere easier to interpret.

Counter-propagating evolution on the Poincare is governed by the Stokes vector equations (c.f. Eqs.~\eqref{eqs:simple_stokes})

\begin{figure}
\centering
	\includegraphics[scale=0.7]{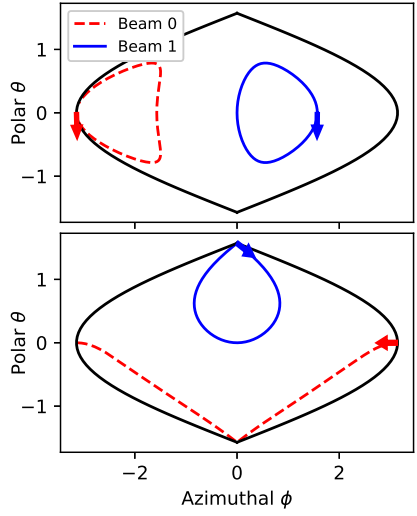}
	\caption{\label{fig:1Dcntr_sphere} Polarization trajectories of two counter-propagating beams displayed on the Stokes's sphere. Arrows indicate initial state and direction of evolution. The polarization state of beam 1 is described from the perspective of beam 0, so a left-circularly polarized beam 1 starts at the north pole (right-circular polarization from the perspective of beam 0). With this convention, the beams trace out parallel circles of equal size. At half a period, the beams reach points antipodal to the other beam's initial state. Beam 0 (red dashed line) is initially s-polarized while beam 1 (blue solid line) is initially either linearly polarized at $45^\circ$ (top) or left-circularly polarized (bottom).}
\end{figure}

\bse\label{eqs:simple_stokes_cntr}
\begin{align}
\parder{}{\zeta}\hat{\bm{J}}_0&=\hat{\bm{J}}_0\times\hat{\bm{J}_1},\label{eq:simple_stokes_cntr_0}\\
\parder{}{\zeta}\hat{\bm{J}}_1&=-\hat{\bm{J}}_1\times\hat{\bm{J}_0},\label{eq:simple_stokes_cntr_1}
\end{align}
\ese
We illustrate two solutions to these equations on the projected Poincare sphere in figure~\ref{fig:1Dcntr_sphere}. The trajectories are parallel circles, with the two beams located at identical points of their respective circles. After a half period of evolution, beam 0 reaches a point on the Poincare sphere antipodal to beam 1's starting location, and similarly for beam 1. Antipodal points on the Poincare sphere correspond to orthogonal polarizations, thus demonstrating the flipping behavior noted in Eq.~\eqref{eq:beam_flip}. The parallel circles are a direct consequence of Eqs.~\eqref{eqs:simple_stokes_cntr}. Subtracting the two equations shows $\hat{\bm{J}}_0-\hat{\bm{J}}_1$ is conserved. This forces the two unit vectors to rotate about an axis given by their difference, causing them to sweep out parallel circles around this axis. After a half period, the vectors return to their initial plane (though not their initial orientations). One can verify that at this point $\hat{\bm{J}}_0(\zeta_\pi/2)=-\hat{\bm{J}}_1(0)$, $\hat{\bm{J}}_1(\zeta_\pi/2)=-\hat{\bm{J}}_0(0)$ because this preserves $\hat{\bm{J}}_0-\hat{\bm{J}}_1$ while placing the vectors in their initial plane. This shows how each vector reaches a point antipodal to the other vector's initial state.

\section{Two-dimensional effects in Nonlinear Polarization Mixing\label{sec:2D}}

In section~\ref{sec:1D} we assumed the electric field amplitudes in Eqs.~\eqref{eqs:full_amplitude} varied only with $z$ and that $\bm{k}_0\propto\bm{k}_1\propto\hat{\bm{z}}$. Now we take the more general case where the amplitudes can vary in the 2D plane spanned by $\bm{k}_0$, $\bm{k}_1$ (the plane of incidence) and the waves propagate along their respective $k$-vectors. We again focus on the case where no energy is exchanged between the beams (extending the formalism to include energy exchange is straightforward). The full 3D solution can then be obtained by stacking the 2D solutions along the direction orthogonal to the plane of incidence. Effectively, this section extends the 2D solution for intensities presented in Ref.~\cite{mckinstrie_1996} to the general case including polarization dynamics. We find the 2D solution has none of the features of the 1D solutions of Sec.~\ref{sec:1D}, making it impossible to use for polarization control except for the special case of $90^\circ$ beam intersection (see Ref.~\cite{michel_PRX2020}). In particular, the transverse beam profiles are highly non-uniform after the interaction with none of the predictable behavior of the 1D solution, even for small intersection angles.

When the beams propagate in different directions, each beam has its own plane for polarization (orthogonal to its direction of propagation). We use the standard $\hat{\bm{s}}_i$, $\hat{\bm{p}}_i$ basis for each beam's polarization where $\hat{\bm{s}}_i=-\hat{\bm{k}}_0\times\hat{\bm{k}}_1/\left|\hat{\bm k}_0\times\hat{\bm{k}}_1\right|$ is orthogonal to the plane of incidence ($s$-polarization) and $\hat{\bm{p}}_i=\hat{\bm{s}}_i\times\hat{\bm{k}}_i$ is in the plane of incidence ($p$-polarization). Note that the $s$-polarization directions coincide for the two beams while the $p$-polarization directions do not. We now use the (un-normalized) Jones vector notation
\beqn
\ket{\mathcal{E}_i}=\colvectwo{{E}_{p,i}}{{E}_{s,i}},\bra{\mathcal{E}_i}=\begin{pmatrix}
	E_{p,i}^* & E_{s,i}^*,
\end{pmatrix}
\eeqn so that $\braket{\mathcal{E}_i}{\mathcal{E}_i}=|\bm{E}_i|^2$. To account for the different orientations of the polarization planes, Eqs.~\eqref{eqs:full_amplitude} require projecting the right-hand side onto the polarization plane of the beam on the left-hand side. When acting on Jones vectors, the projection operators take the form
\beqn
P_{01}=P_{10}=\begin{pmatrix} \cos\psi & 0\\ 0 & 1\end{pmatrix},
\eeqn where $\psi$ is the angle of intersection between the beams.

We define (skew) coordinates $\zeta_0,\zeta_1$ so that the position vector is $\bm{r}=\zeta_0\hat{\bm{k}}_0+\zeta_1\hat{\bm{k}}_1$. For convenience we place one of the edges of beam 0 at $\zeta_1=0$ and similarly for beam 1 so $(0,0)$ occurs at the corner of the parallelogram formed by the intersection of the beams (for definiteness we take this to be the corner closest to the laser sources). With these definitions, Eqs.~\eqref{eqs:full_amplitude} become
\bse\label{eqs:dirac_amplitude}
\begin{align}
\parder{}{\zeta_0}\ket{\mathcal{E}_0} &=i\frac{n_2k_0c}{8\pi}P_{01}\ket{\mathcal{E}_1}\bracket{\mathcal{E}_1}{P_{10}}{\mathcal{E}_0},\label{eq:dirac_amplitude_0}\\
\parder{}{\zeta_1}\ket{\mathcal{E}_1} &=i\frac{n_2k_1c}{8\pi}P_{10}\ket{\mathcal{E}_0}\bracket{\mathcal{E}_0}{P_{01}}{\mathcal{E}_1}.\label{eq:dirac_amplitude_1}
\end{align}
\ese

To solve these equations we specify initial conditions $\ket{\mathcal{E}_0(\zeta_0=0,\zeta_1)}$ and $\ket{\mathcal{E}_1(\zeta_0,\zeta_1=0)}$. We can then integrate from the corner $(0,0)$ using, for example,
\beqn
\ket{\mathcal{E}_0(\zeta_0+\Delta\zeta_0,\zeta_1=0)}=\ket{\mathcal{E}_0(\zeta_0,\zeta_1=0)}
+\Delta\zeta\left(i\frac{n_2k_0c}{8\pi}\right)P_{01}\ket{\mathcal{E}_1}\bracket{\mathcal{E}_1}{P_{10}}{\mathcal{E}_0}\bigg|_{(\zeta_0,\zeta_1=0)},
\eeqn and repeating along the $\zeta_1=0$ line (where $\ket{\mathcal{E}_1}$ is known). Having thus determined $\ket{\mathcal{E}_0(\zeta_0,\zeta_1=0)}$ we can integrate beam 1 according to
\beqn
\ket{\mathcal{E}_1(\zeta_0,\zeta_1=\Delta\zeta)}=\ket{\mathcal{E}_1(\zeta_0,\zeta_1=0)}
+\Delta\zeta\left(i\frac{n_2k_1c}{8\pi}\right)P_{10}\ket{\mathcal{E}_0}\bracket{\mathcal{E}_0}{P_{01}}{\mathcal{E}_1}\bigg|_{(\zeta_0,\zeta_1=0)}.
\eeqn In short, by using these integration steps we are ensuring the right-hand side is known when the step takes place. We repeat the above integration steps to propagate the solution through the entire interaction region.

\begin{figure}
\centering
	\includegraphics[scale=0.7]{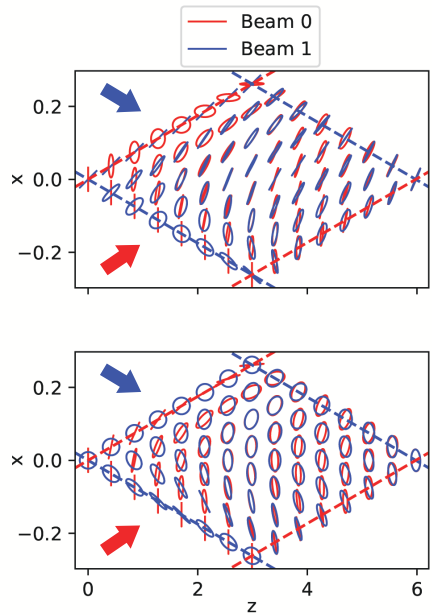}
	\caption{\label{fig:2DwoCBET} Polarization dynamics of two beams intersecting at $10^\circ$ (aspect ratio changed for display purposes), obtained by solving Eqs.~\eqref{eqs:dirac_amplitude}. Beams propagate towards positive $z$ (as indicated by the arrows) with beam 0 initially s-polarized and beam 1 initially either linearly polarized at $45^\circ$ (top) or right-circularly polarized (bottom). Strong transverse non-uniformity of polarization develops as different parts of the beams see different polarization states of the other beam.}
\end{figure}

We plot solutions to Eqs.~\eqref{eqs:dirac_amplitude} in Fig.~\ref{fig:2DwoCBET}, with beam 0 initially $s$-polarized and beam 1 initially either polarized at $45^\circ$ from its $p$-direction or right-circularly polarized. We use the normalized coordinates 
\bse
\begin{align}
x &=\frac{n_2kc}{8\pi}(\zeta_0-\zeta_1)\sin(\psi/2)\\
z &=\frac{n_2kc}{8\pi}(\zeta_0+\zeta_1)\cos(\psi/2),
\end{align}
\ese where $k=k_0=k_1$. We see strong transverse non-uniformity develop as the beams evolve, with none of the 1D polarization swapping dynamics. This is caused by different parts of the beam experiencing different conditions of the other beam. For example, the left edge of beam 0 sees a single, fixed polarization of beam 1 causing beam 0's polarization state to evolve along the edge. Different parts of beam 1 now interact with different polarization states of beam 0 leading to transverse non-uniformity which in turn impacts the rest of beam 0. Furthermore, unlike the 1D case, each beam constantly encounters new rays of the other beam that it has not interacted with yet, breaking the feedback present in the 1D solution.

\section{Recovering One-Dimensional Solutions with a Thin Nonlinear Medium\label{sec:ThinPlasma}}
\begin{figure}
\centering
	\includegraphics[scale=0.7]{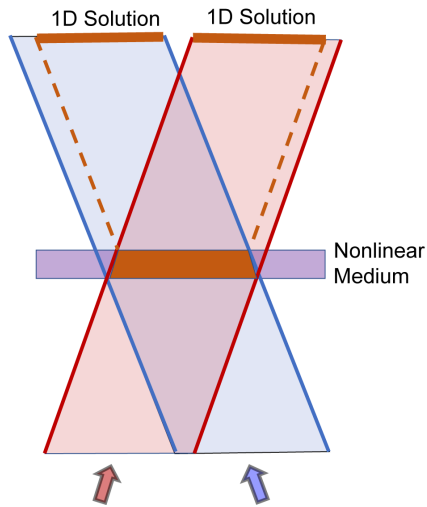}
	\caption{\label{fig:2D1D_edge} Illustration of how the edge effects impinge on the 1D solution region. When two uniform beams reach the boundary of a nonlinear medium, translation invariance guarantees a 1D solution except at the edges of the beams. These edge disturbances propagate along each beam's rays narrowing the region that supports the 1D solution in both beams.}
\end{figure}

In the previous section we showed that 2D dynamics causes strong deviation from the 1D model predictions. In order to reproduce the 1D results, we reduce the nonlinear medium to a thin strip near the widest part of the beam intersection region (see Fig.~\ref{fig:2D1D_edge}). This allows the central parts of both beams to follow the 1D solution. For small intersection angles, we thus restore polarization swapping allowing us to control the beam polarization using a secondary beam as in Sec.\ref{subsec:coprop}.

By introducing a thin nonlinear medium as in Fig.~\ref{fig:2D1D_edge}, we allow the beams to have significant overlap before they begin interacting. The interaction is still governed by Eqs.~\eqref{eqs:dirac_amplitude}, but now the right-hand side is constant along the initial boundary of the nonlinear medium (assuming initially uniform beams) except for at the beam edges. This translational invariance (parallel to the medium boundary) is preserved by Eqs.~\eqref{eqs:dirac_amplitude}, meaning the beam amplitudes only vary along the dimension orthogonal to the boundary. This brings us back to the 1D model of Sec.~\ref{sec:1D} with $k_z$ replacing $k$ for each beam (to account for the difference between $\zeta_0,\zeta_1$ of Sec.~\ref{sec:2D} and $\zeta$ of Sec.~\ref{sec:1D}). However, the translational invariance is broken at the edges of the beams. This edge disturbance propagates along the rays of the two beams, impinging on the region supporting the 1D solution (see Fig.~\ref{fig:2D1D_edge}). 

A trapezoidal region of 1D dynamics is thus formed: the base is given by the beam overlap at the front boundary of the nonlinear medium, the sides are given by the rays of the beams, and the top is then determined by the location of the rear boundary of the nonlinear medium. The ratio of the top of the trapezoid to the beam width determines the fraction of the beam that obeys the 1D solution. Increasing the thickness of the nonlinear medium decreases the size of the top of the trapezoid, thus limiting the fraction of each beam obeying the 1D solution. This occurs whether we bring the front boundary closer (thus decreasing the base of the trapezoid and indirectly shrinking its top) or move the rear boundary further (directly reducing the size of the top.) Thus we need to keep the nonlinear boundary thin compared to the beam overlap volume as illustrated in Fig.~\ref{fig:2D1D_edge}.

\begin{figure}
\centering
	\includegraphics[scale=0.7]{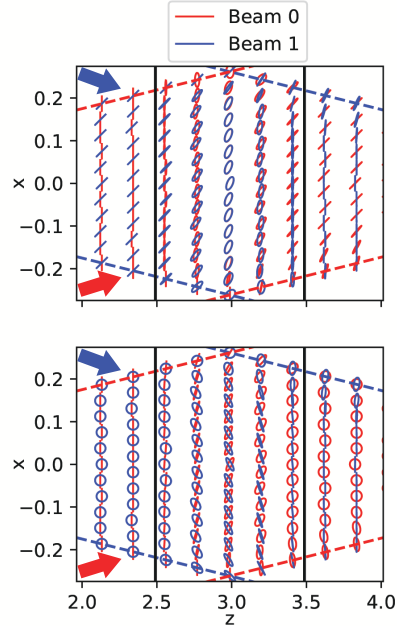}
	\caption{\label{fig:ThinPlasma}Similar to Fig.~\ref{fig:2DwoCBET} in the presence of a thin nonlinear medium (boundaries of the medium indicated by vertical black lines), zoomed in to show the interaction region. The central regions of both beams follow the 1D solution, demonstrating polarization swapping. Deviation from the 1D solution occurs near the edges of both beams, as expected.}
\end{figure}

We demonstrate how using this thin nonlinear medium produces the 1D solution in Fig.~\ref{fig:ThinPlasma}. The parameters are identical to Fig.~\ref{fig:2DwoCBET} with the only change being the extent of the nonlinear medium (its boundaries are indicated by solid black lines in Fig.~\ref{fig:ThinPlasma}). We can now see the polarization swapping phenomenon occurring in the central regions of both beams, with deviations near the edges of the beams as expected. We can now utilize the polarization control described in Sec.~\ref{sec:1D} within a 2D propagation geometry.

\section{Conclusion\label{sec:conclusion}}
We have presented schemes for polarization control of equal intensity, equal wavelength laser beams using both co- and counter-propagating 1D geometries. The schemes consisted of polarization swapping (co-propagating) and flipping (counter-propagating) between the beams. We demonstrated how 2D geometries scramble the 1D results and how to mitigate these effects by using a thin nonlinear medium. We now consider some applications of our schemes for optical computing and cryptography.

In optical computing, we can create a NOT gate in a two-stage process using both our swapping and flipping schemes. First, note that we can flip a polarization of a beam to its orthogonal state. Suppose beam 0 is initially vertically polarized and interacts with a circularly polarized beam 1 in the swapping (co-propagating) geometry. After this first interaction (with beam 1 now vertically polarized), we can counter-propagate the two beams (by, for example, reflecting beam 1) causing beam 0 to flip to a horizontal polarization, orthogonal to its initial state. The same setup will work to change a horizontal polarization to a vertical one. If we use orthogonal polarizations (e.g. horizontal and vertical) to encode 0- and 1-bits, then the setup just described is a NOT gate taking 0 to 1 and vice versa.

For cryptography, we take advantage of the transformations of the Poincare sphere our polarization control schemes induce. If we fix the polarization state of one of the beams (the auxiliary), then each polarization state of the other beam (the probe) is mapped to a new one (in either the co- or counter-propagating geometry), creating a map from the Poincare sphere to itself. Here we want to avoid using the swap or flip distance as then the map becomes singular. If we use the northern and southern hemispheres of the Poincare sphere to represent 0- and 1-bits respectively, then we can use the map of the Poincare sphere to encrypt 0-bits by sampling probe beam polarization states from the inverse image of the northern hemisphere (and similarly for 1-bits). Having the probe beam then interact with the appropriate auxiliary beam will decrypt the message.

We thus see how our polarization control schemes can be applied in a wide variety of applications, opening the door for novel methods of laser beam control. 

\section*{Acknowledgments}
The authors would like to thank Andrew Charman (UC Berkeley) and Lazar Friedland (Hebrew University of Jerusalem) for helpful discussions. This work was performed under the auspices of the U.S. Department of Energy by Lawrence Livermore National Laboratory under contract DE-AC52-07NA27344 the NSF-DOE Partnership in Plasma Science under Grant 1803874, and the LLNL-LDRD Program under Project tracking No. 18-ERD-046.

\section*{Disclosures}
The authors declare no conflicts of interest.

\bibliography{CBET_and_Polarization_Dynamics_Paper}






\end{document}